\documentclass[manuscript]{aastex63}

\newcommand{\rsun}{$R_\sun$}
\newcommand{\kms}{km s$^{-1}$}

\usepackage{bm}
\usepackage{graphicx}
\usepackage{subfigure}
\begin{document}
	
\title{Multispacecraft Remote Sensing and In Situ Observations of the 2020 November 29 Coronal Mass Ejection and Associated Shock: From Solar Source to Heliospheric Impacts}

\author{Chong Chen}
\affiliation{State Key Laboratory of Space Weather,	National Space Science Center, Chinese Academy of Sciences, Beijing 100190, People’s Republic of China; \href{mailto:liuxying@swl.ac.cn}{liuxying@swl.ac.cn}}
\affiliation{University of Chinese Academy of Sciences, Beijing 100049, People’s Republic of China}

\author{Ying D. Liu}
\affiliation{State Key Laboratory of Space Weather,	National Space Science Center, Chinese Academy of Sciences, Beijing 100190, People’s Republic of China; \href{mailto:liuxying@swl.ac.cn}{liuxying@swl.ac.cn}}
\affiliation{University of Chinese Academy of Sciences, Beijing 100049, People’s Republic of China}

\author{Bei Zhu}
\affiliation{Space Engineering University, Beijing 101416, People’s Republic of China}

\begin{abstract}
We investigate the source eruption, propagation and expansion characteristics, and heliospheric impacts of the 2020 November 29 coronal mass ejection (CME) and associated shock, using remote sensing and in situ observations from multiple spacecraft. A potential--field source--surface model is employed to examine the coronal magnetic fields surrounding the source region. The CME and associated shock are tracked from the early stage to the outer corona using extreme ultraviolet and white light observations. Forward models are applied to determine the structures and kinematics of the CME and the shock near the Sun. The shock shows an ellipsoidal structure, expands in all directions, and encloses the whole Sun as viewed from both SOHO and STEREO A, which results from the large expansion of the CME flux rope and its fast acceleration. The structure and potential impacts of the shock are mainly determined by its radial and lateral expansions. The CME and shock arrive at Parker Solar Probe and STEREO A. Only based on the remote sensing observations, it is difficult to predict whether and when the CME/shock would arrive at the Earth. Combining Wind in situ measurements and WSA-ENLIL simulation results, we confirm that the far flank of the CME (or the CME leg) arrives at the Earth with no shock signature. These results highlight the importance of multipoint remote sensing and in situ observations for determining the heliospheric impacts of CMEs.

\end{abstract}


\section{Introduction} \label{sec:intro}
Coronal mass ejections (CMEs) are large--scale magnetized plasma ejected from the Sun into interplanetary space, which are responsible for major geomagnetic storms in the terrestrial environment. They are called interplanetary CMEs (ICMEs), when traveling into interplanetary space. A fast CME can drive a shock ahead of it, when the CME speed relative to the ambient solar wind is greater than the magnetosonic speed or Alfv\'en speed of the ambient solar wind. CME--driven shocks can accelerate solar energetic particles (SEPs) and enhance the geo--effectiveness of CMEs. Understanding the process of CME/shock propagation and evolution is of critical importance for space weather forecasting. 

Multispacecraft remote sensing and in situ observations are required to study the complete chain of the evolution of CMEs/shocks from their solar sources through heliospheric propagation to their impacts. As far as we know, the study of such a complete chain is still lacking. 
First, the use of multi--perspective remote sensing observations can better constrain the three--dimensional (3D) structure of the CME and its associated shock. To determine the 3D morphology of a CME, \citet{Thernisien2006,Thernisien2009} develop a graduated cylindrical shell (GCS) model based on the coronagraph images from different vantage points. As for CME--driven shocks, \citet{Kwon2014, Kwon2015} propose an ellipsoid model to reconstruct the 3D structure of a shock. 
Second, separated multipoint in situ measurements can better assess the structure and the heliospheric impacts of CMEs and shocks (e.g., \citealt{Liu2008, Liu2019, Hu2017, Lucas2011}). The multispacecraft measurements at different locations can determine the propagation and expansion characteristics of CMEs/shocks along different directions, including the evolution anisotropy of CMEs/shocks resulting from different background solar wind. Multipoint in situ measurements are also needed to verify the forecasting accuracy along different directions of a magnetohydrodynamics model in a same event simulation (e.g., \citealt{Reinard2012,Biondo2021}).

Finally, the combination of remote sensing and in situ observations from multiple spacecraft can provide a more complete picture of CME/shock evolution, and improve our capabilities to interpret events and to forecast space weather effects (e.g., \citealt{Liu2008,Liu2011,Liu2012, Nieves2012,Hu2016,Palmerio2021}). 
Below we show examples of how the multi--perspective remote sensing observations in combination with in situ measurements are used to investigate the 3D propagation and evolution of CMEs/shocks. 
Based on the wide--angle imaging observations from Solar Terrestrial Relations Observatory (STEREO; \citealt{Kaiser2008}) and in situ measurements at 1 AU, a geometric triangulation technique has been developed to track CMEs \citep{Liu2010a, Liu2010b}, and the Sun--to--Earth kinematics of fast and slow CMEs have been revealed \citep{Liu2013, Liu2016}. 
Analyzing multipoint imaging observations and widely separated in situ measurements, \citet{Liu2017, Liu2019} find the fading and persistence of CME--driven shocks along different directions in the heliosphere. 
In addition, using multipoint remote sensing and in situ observations, the interactions of successive CMEs, which may cause intense geomagnetic storms, are examined during the whole propagation process (e.g., \citealt{Liu2012,Liu2014NatCo,Liu2014ApJ,Webb2013,Mishra2015,Lugaz2017}).
The geometrical and magnetic relationships between ICMEs at 1 AU, CMEs near the Sun and their solar sources have also been investigated, using coordinated imaging and in situ observations from multiple vantage points (e.g., \citealt{Liu2010b, Liu2011, Xie2021,Marubashi2015,Syed2019,Savani2015}). However, the connections between in situ signatures and solar source characteristics are still not fully understood, because CMEs can have a very complex evolution from the Sun to 1 AU, including deceleration, deflection, rotation, erosion, and interaction with ambient solar wind structures and other CMEs (e.g., \citealt{2017Manchester} and references therein). Therefore, the comparisons between multi--perspective remote sensing observations and multipoint in situ measurements are necessary and critical to make the correct connections between the solar source eruptions, CMEs near the Sun and their in situ counterparts and to understand their heliospheric impacts.
 
\begin{figure}
	\epsscale{0.8}
	\plotone{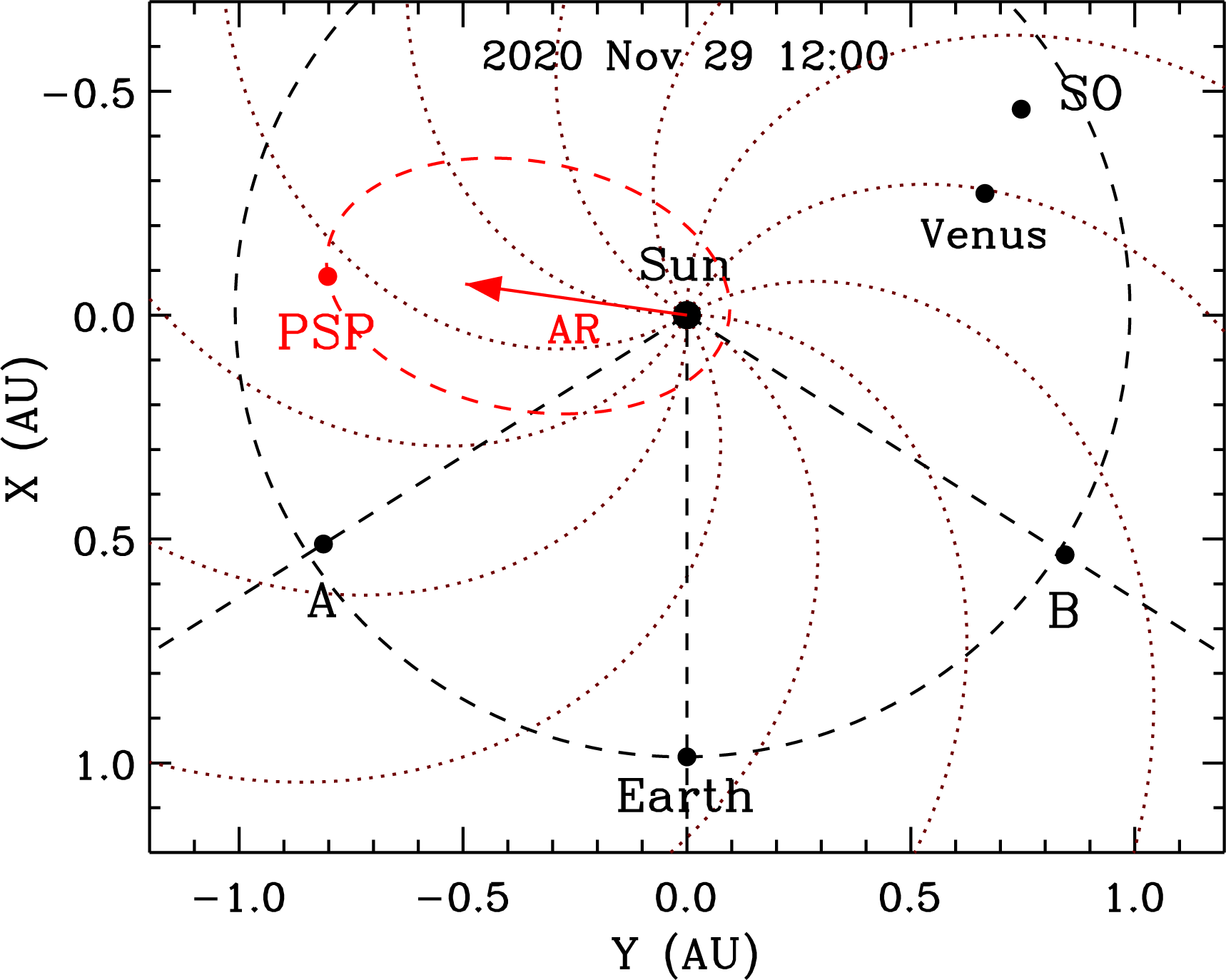}
	\caption{\label{f1}Positions of the spacecraft and planets in the ecliptic plane on 2020 November 29. The ``A'', ``B'', and ``SO'' represent the satellites of STEREO A, STEREO B, and Solar Orbiter respectively. The black circle and red ellipse mark the orbits of the Earth and PSP respectively. The dotted curves show Parker spiral magnetic fields created with a solar wind speed of 450 \kms{}. The red arrow indicates the direction of the source region in the ecliptic plane.} 
\end{figure}

On 2020 November 29, a large CME erupted, which is associated with a shock and an M4.4 class flare that peaked at about 13:11 UT. The CME caused the first type II radio burst and the first widespread SEP event of solar cycle 25 \citep{Lario2021, Kollhoff2021}. The eruption is a limb event as viewed from the Earth, and is observed by a fleet of spacecraft at different vantage points. The positions of the spacecraft in the ecliptic plane on 2020 November 29 are shown in Figure \ref{f1}. STEREO A and Parker Solar Probe (PSP; \citealt{Fox2016}) are 0.96 AU and 0.81 AU from the Sun, and 57.8\degr{} and 96.2\degr{} east of the Earth, respectively. The red arrow indicates the location of the source region (E97\degr{}S25\degr{}), which points to PSP in the ecliptic plane. 
This event provides a unique opportunity to investigate the solar source, propagation, and heliospheric impacts of the CME and its shock, as multi--perspective remote sensing observations and multipoint in situ measurements are available for this event.
We examine source eruption signatures in Section \ref{source}, the propagation and expansion characteristics of the CME/shock in Section \ref{evolution}, and in situ measurements at different locations in Section \ref{insitu}. The results are summarized in Section \ref{sum}. The results of this work provide a more complete propagation and expansion picture of the CME and its shock, and highlight the importance of multispacecraft remote sensing and in situ observations for determining the heliospheric impacts of CMEs.

\section{source eruption signatures} \label{source}
The CME was launched around 12:45 UT on 2020 November 29 associated with significant features, such as an M4.4 class flare, coronal dimmings, and EUV waves as shown in Figure \ref{f2}. Panel (a) shows an overview of the source region in a
195 \AA{} image from the Extreme--Ultraviolet Imager (EUVI) of the Sun Earth Connection Coronal and Heliospheric Investigation (SECCHI; \citealt{Howard2008}) aboard STEREO A with potential--field source--surface (PFSS) extrapolation results mapped on it. The viewpoint of the image deviates a few degrees from STEREO A (see the black part on the left of the EUVI image) to show the coronal magnetic field configuration around the source region. The closed magnetic field arches are toroidal and surrounded by open magnetic field lines of the same polarity. The source region is beneath the right part of the toroidal closed magnetic field arches as marked by the white arrow in panel (a). 

Panels (b)-(c) display STEREO A/EUVI 195 \AA{} running--difference images at two moments during the time of the eruption. The solar flare is visible at the beginning of the eruption followed by a significant dimming region, which is indicative of the removal of the coronal plasma. The dimming region continuously extends. Panel (b) also shows the significant rise and expansion of the plasma loop. The most dramatic signature in the EUV observations is the EUV wave around the dimming, which propagates away from the source region and sweeps a large area of the solar disk, as shown in panel (c). The continuously extended areas of the dimming and the EUV wave indicate the expansion of the CME as it rises up. 

Panel (d) shows a hot channel at its nascent stage, which has been interpreted as a flux rope (e.g., \citealt{ZhangJ2012,ChengX2013,ChengX2014}) although the magnetic field is not observed, in a 131 \AA{} image observed by the Atmosphere Imaging Assembly (AIA; \citealt{Lemen2012}) on board Solar Dynamics Observatory (SDO; \citealt{Pesnell2012}). This is a limb event as viewed from the Earth, and the source region is just behind the solar limb, so we can track the hot channel at the early stage using EUV observations (see below). The axis of the hot channel possibly has a largely inclined angle, because the two legs of the hot channel separate with a large distance. SDO/AIA 193 \AA{} running--difference images are displayed in panels (e)-(f). In panel (e), the plasma loop is bubble--shaped ahead of the hot channel shown in panel (d). There is an ejection propagating toward the south as marked by the red arrow in panel (f). This is not the eruption we study, because of its propagation direction. This eruption may be caused by the EUV wave, which sweeps across and destabilizes it.

These observations suggest that the hot channel has a large tilt angle, and, as the hot channel rises and expands, it causes a dimming region and a large--scale EUV wave. The EUV wave propagates across a significant portion of the solar disk and is recognized as the footprint of the shock (see below).

\section{structure and kinematics of the CME/shock} \label{evolution}
\begin{figure}
	\epsscale{1.1}
	\plotone{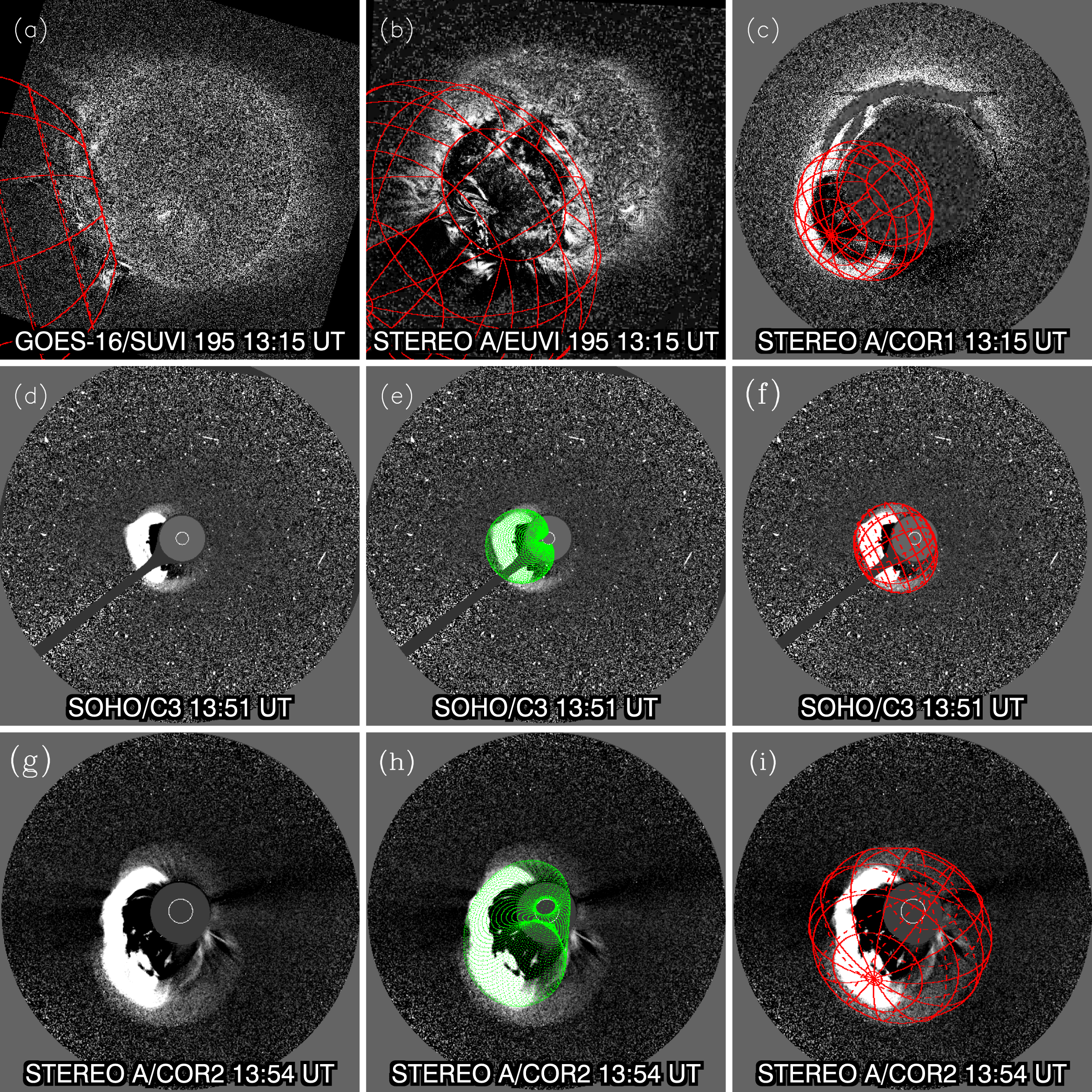}
	\caption{\label{f3}Running--difference images from GOES 16, STEREO A, and SOHO with corresponding modeling of the CME and shock. Panels (a)-(c) show the same shock modeling result for 13:15 UT. Panels (d)-(i) display the running--difference coronagraph images and corresponding GCS modeling (green) and ellipsoid shock modeling (red).}
\end{figure}

EUV and coronagraph observations of the eruption from GOES 16, STEREO A, and Solar and Heliospheric Observatory (SOHO; \citealt{Domingo1995}) are displayed in Figure \ref{f3}. Panels (a)-(c) show the eruption at the early stage in EUV and white light images. The primary EUV wave is considered to be the footprint of the CME--driven shock (e.g., \citealt{Patsourakos2012, Cheng2012, Kwon2014}), and at the early time the fronts of the CME and shock are not well separated (see panel (c)). Therefore, they can be used together to determine the shock structure. The well--developed CME and shock in coronagraph images from two vantage points are shown in panels (d) and (g). The shock can be seen in the images as a faint edge around the CME leading edge (e.g., \citealt{Liu2008, Hess2014,Zhu2021}). We can see the backward propagation of the shock on the opposite side of the eruption, because it expands in all directions. Previous studies suggest that the shock can be modeled well using a simple spheroidal structure (e.g., \citealt{Hess2014, Kwon2014, Liu2017, Liu2019}). Here, we use a geometrical ellipsoid model developed by \citet{Kwon2014} to determine the 3D morphology of the shock, based on the EUV and white light observations from GOES 16, STEREO A, and SOHO. The ellipsoid model has seven free parameters: the height, longitude, and latitude of the center of the ellipsoid; the lengths of the three semiprincipal axes; and the rotation angle of the ellipsoid. We assume the cross section of the shock ellipsoid perpendicular to the propagation direction to be a circle, which reduces two free parameters (one semiprincipal axis and the rotation angle) in our fitting. We start to fit the shock at 12:56 UT when the EUV wave can be distinguished from the brightenings in the dimming region, and get a good visual consistency between the model and observations. At the initial stage, the footprint of the ellipsoid model on the solar disk is consistent with the EUV wave front (see panels (a)-(b)). In coronagraph images, as shown in panels (c), (f), and (i), the shock front is well represented by the ellipsoid model. As for the CME, we employ the GCS model proposed by \citet{Thernisien2006} to fit the CME based on running--difference coronagraph images from STEREO A and SOHO. The GCS model can determine the direction of propagation, tilt angle of CME flux rope and height. By adjusting the free parameters, the GCS model fits the CME observations from the two viewpoints very well, as shown in panels (e) and (h). 

\begin{figure}
	\epsscale{1.2}
	\plotone{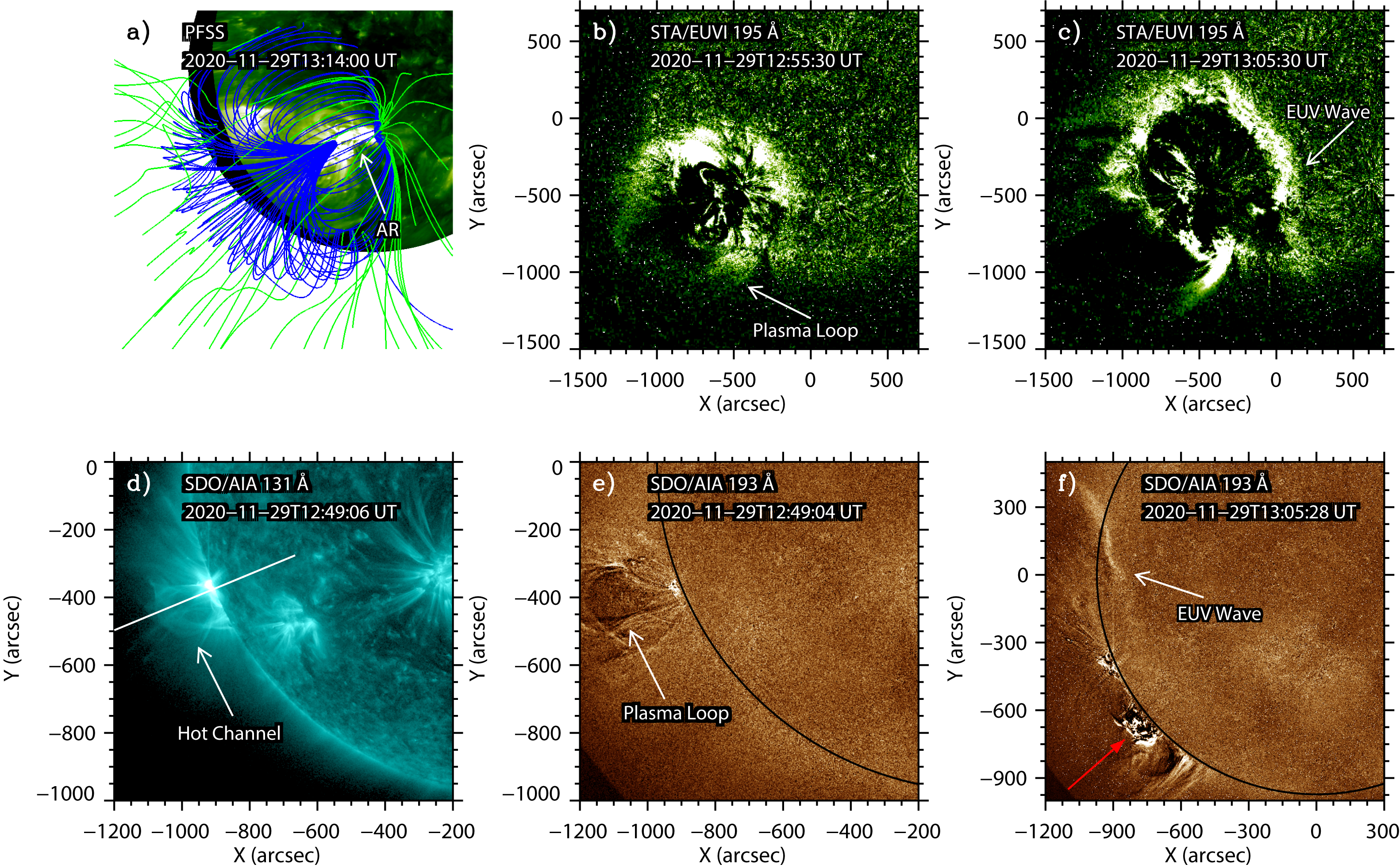} 
	\caption{\label{f2}Magnetic field configuration and EUV observations of the source region. (a) PFSS modeled coronal magnetic fields surrounding the source region mapped onto the EUVI 195 \AA{} image at 13:14 UT on 2020 November 29. The blue lines represent the closed field lines, and the green lines are the open magnetic field lines. The white arrow marks the source region. (b)-(c) Running--difference images of STEREO A/EUVI at 195 \AA{} showing the low coronal signatures associated with the CME. (d) Hot channel at 131 \AA{} from SDO/AIA. The white line indicates a slice along the radial direction to create a distance–time diagram for the hot channel. (e)-(f) Running--difference images of SDO/AIA at 193 \AA{} showing the plasma loop and the EUV wave from the viewpoint of the Earth. The red arrow marks the eruption caused by the EUV wave.} 
\end{figure}

\begin{figure}
	\epsscale{1}
	\plotone{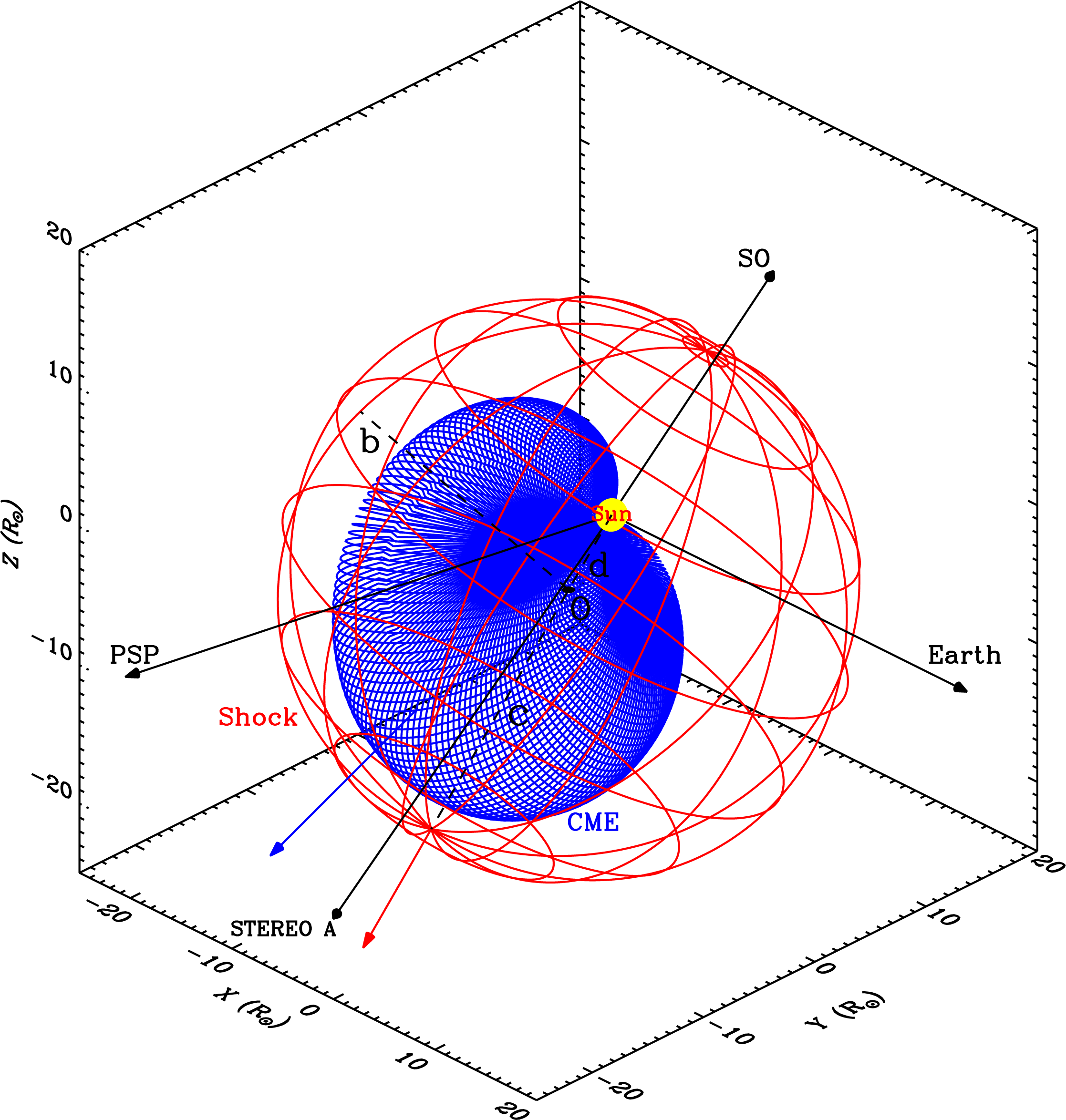}
	\caption{\label{f4}Geometry of the CME (blue lines) and the shock (red lines) at 15:18 UT on 2020 November 29. The blue and red arrows indicate the propagation directions of the CME and shock, respectively. The directions of the Earth, STEREO A, PSP, and Solar Orbiter are marked by the black arrows. The center of the shock ellipsoid (O) and its distances from the center of the Sun (d), from the nose of the shock (c) and from the flank perpendicular to the propagation direction (b) are also indicated.
} 
\end{figure}

Figure \ref{f4} shows the 3D modeled CME and shock structures at 15:18 UT on 2020 November 29. At this moment, the propagation longitude of the CME and shock is about 85\degr{} and 80\degr{} east of the Sun–Earth line as marked by the blue and red arrows, respectively. Application of the GCS model gives an average propagation direction of about 90\degr{} east of the Sun–Earth line and about 17\degr{} south of the ecliptic plane. The propagation direction of the CME is roughly consistent with the location of the source region (E97\degr{}S25\degr{}), and does not change much within 30 \rsun{}. The tilt angle of the CME flux rope obtained from the GCS model is about 75\degr{} with respect to the ecliptic plane, which is consistent with the AIA 131 \AA{} observations (see Figure \ref{f2} (d)). When we employ the GCS model to fit the CME (e.g., \citealt{Thernisien2006, Thernisien2009, Liu2010b}), the half angle increases from 15\degr{} at 2.4 \rsun{} to 70\degr{} at 15.8 \rsun{}, demonstrating the large expansion of the flux rope at its initial stage. The speed of the CME leading edge is accelerated from $\sim$800 \kms{} to $\sim$2100 \kms{} within 10 \rsun{} (see below). Therefore, the ellipsoidal structure of the shock is produced by the large expansion of the CME flux rope and its fast acceleration. The front of the shock nose is close to the CME nose, while the flank has farther distances from the CME. The shock surrounds the whole Sun, which helps to explain the detection of SEPs at Solar Orbiter behind the location of the eruption (see Figure \ref{f1}). The CME and shock could arrive at PSP and STEREO A, according to their propagation directions and large expansions.  

\begin{figure}
	\epsscale{1.3}
	\plotone{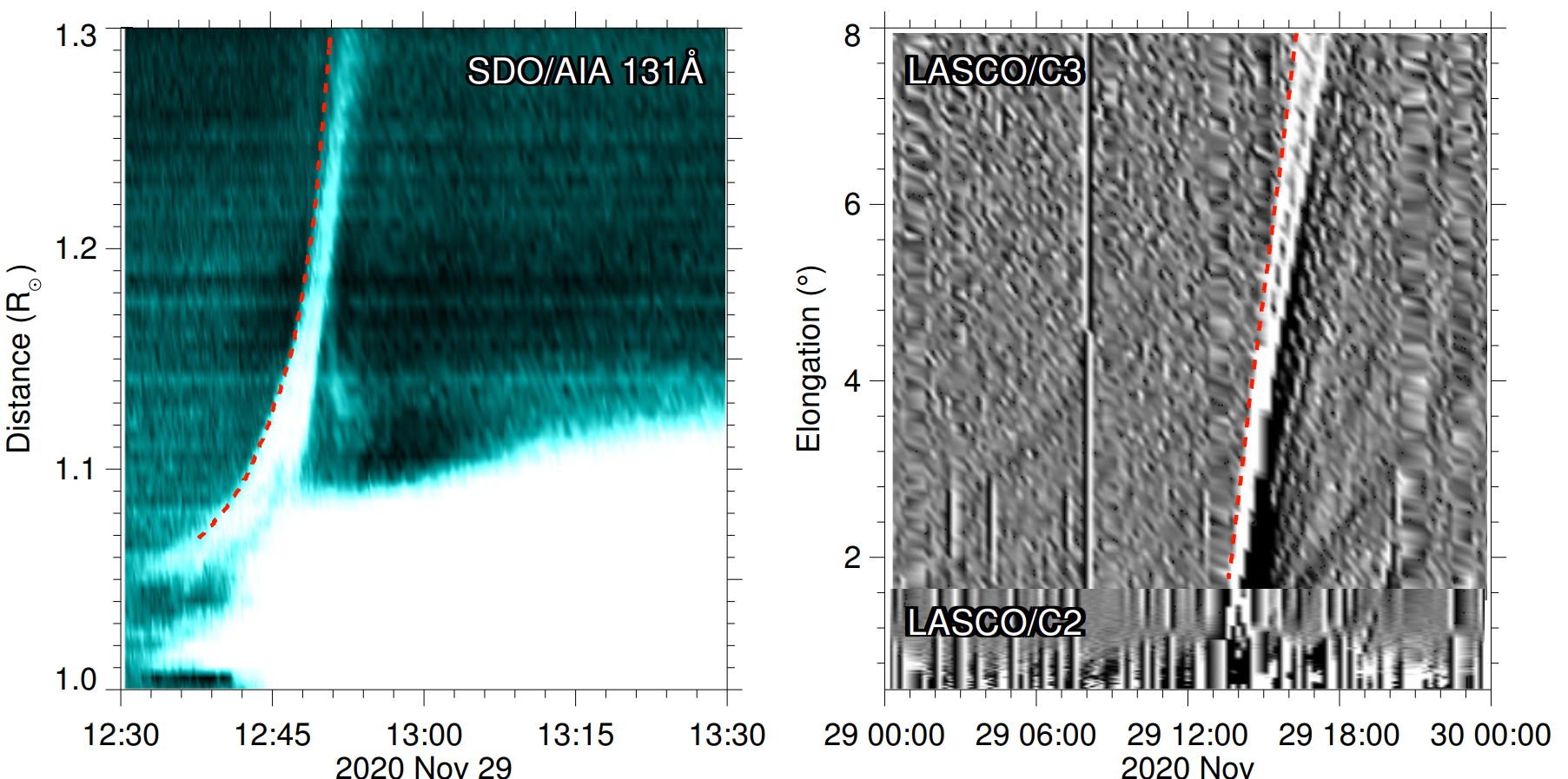}
	\caption{\label{f5}Left: distance--time map of the hot channel by stacking SDO/AIA 131 \AA{} running--difference images along the white line in the panel (d) of Figure \ref{f2}. The red curve indicates the track of the hot channel. Right: time--elongation map constructed from running--difference images of C2 and C3 from SOHO/LASCO along the ecliptic plane. The red curve marks the track of the CME leading edge, along which the elongation angles are extracted.} 
\end{figure}

We can track the CME flux rope and the shock from their initial stage to the outer corona with projection effects minimized, as the propagation direction is almost perpendicular to the Sun--Earth line. The translation and expansion distances of the shock can be obtained from the ellipsoid model fitting. As shown in Figure \ref{f4}, the lateral and radial expansion distances are represented by ``b'' and ``c'', respectively. The ``d'' and ``d + c'' denote the translation distance of the shock center and shock nose height, respectively. The motion of the CME flux rope can be tracked by stacking the EUV running--difference images within a slit along the radial direction, as marked by the white line in the panel (d) of Figure \ref{f2}. The distance--time map of 131 \AA{} running--difference images from SDO/AIA is shown in the left panel of Figure \ref{f5}. The heights of the hot channel are extracted along the track, as indicated by the red dashed curve. The right panel of Figure \ref{f5} shows the time--elongation map produced by stacking running--difference images of C2 and C3 from SOHO/LASCO within a slit along the ecliptic plane. The track associated with the CME is marked by the red dashed curve. The track in C2 is not clear, because the CME is so fast that C2 only captures three images. Elongation angles of the CME leading edge in the ecliptic plane are extracted along the track. Given that SOHO is about 90\degr{} away from the propagation direction of the CME, we use a harmonic mean (HM) approximation \citep{Lugaz2009}, which assumes that CMEs are attached to the Sun as a spherical front and move along a fixed radial direction, to derive the CME kinematics. Readers are directed to \citet{Liu2013} for discussions of selection of CME geometry depending on the observation angle of the spacecraft.

\begin{figure}
	\epsscale{0.9}
	\plotone{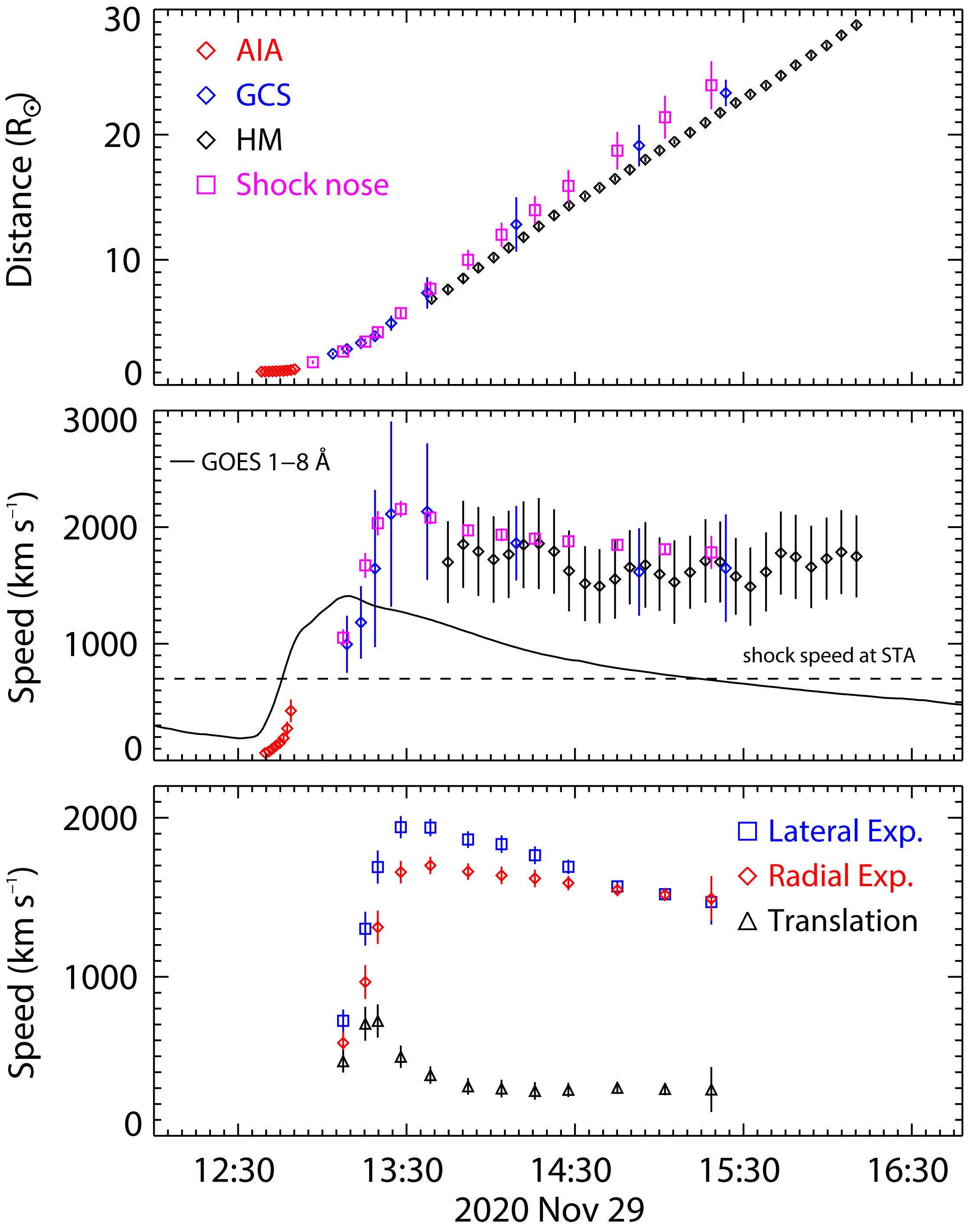}
	\caption{\label{f6}Kinematics of the CME and shock derived from different methods. Top: the distances of the CME derived from EUV observations, the GCS model, and the HM approximation, and the height of the shock nose derived from the ellipsoid model. Middle: speeds of the CME and shock nose derived from the numerical differentiation of the distances in the top panel. GOES 1--8 \AA{} X--ray flux is overlapped in the panel scaled in arbitrary units. The horizontal dashed line represents the shock speed of $\sim$700 \kms{} at STEREO A. Bottom: expansion speeds of the shock along the radial and lateral directions and translation speed of the shock center. The distances and speeds of the CME front are binned to reduce the uncertainties, so the CME parameters are average values and standard deviations within the bins. Following \citet{Kwon2014}, the uncertainty of the ellipsoid model parameters is estimated to be 8\%.} 
\end{figure}

The height and speed of the CME obtained from the GCS modeling, HM approximation, and EUV observations, and the kinematics of the shock are displayed in Figure \ref{f6}. Note that we have used the GCS propagation longitude (E90\degr{}) as input for the HM approximation. As shown in the top panel, the CME height from the HM approximation can connect with the GCS model. The height of the shock nose is slightly larger than the CME height, because the front of the shock nose is ahead of and close to the CME leading edge (see Figure \ref{f4}). The radial distance of the CME flux rope obtained from the EUV observations could be connected with the height of the CME from the GCS model. The speeds of the CME and shock nose are shown in the middle panel. The speed profiles of the CME and shock nose are similar. The CME and shock nose are accelerated to about 2100 \kms{} within only 1 hour below $\sim$6 \rsun{}. Then the speeds have a deceleration. According to the propagation characteristics of fast CMEs described by \citet{Liu2013}, the speeds probably have a gradual longer deceleration in interplanetary space to meet the shock speed at STEREO A (about 700 \kms), as marked by a horizontal dashed line in the middle panel. The bottom panel shows the speeds of the lateral expansion, radial expansion and translation of the shock. The expansion speeds of the shock are much larger than the translation speed of the shock center. Therefore, the shock radial expansion provides a major contribution to the propagation of the shock nose, and the structure and potential impacts of the shock are mainly determined by its radial and lateral expansions. We also overlap the GOES 1--8 \AA{} X--ray flux on the middle panel scaled in arbitrary units. The liftoff of the CME flux rope and the rise of the X--ray flux are almost at the same time. All the speeds increase during the flare rising phase, and reach maxima after the peak value of the X--ray flux. 

\begin{figure}
	\epsscale{1.2}
	\plotone{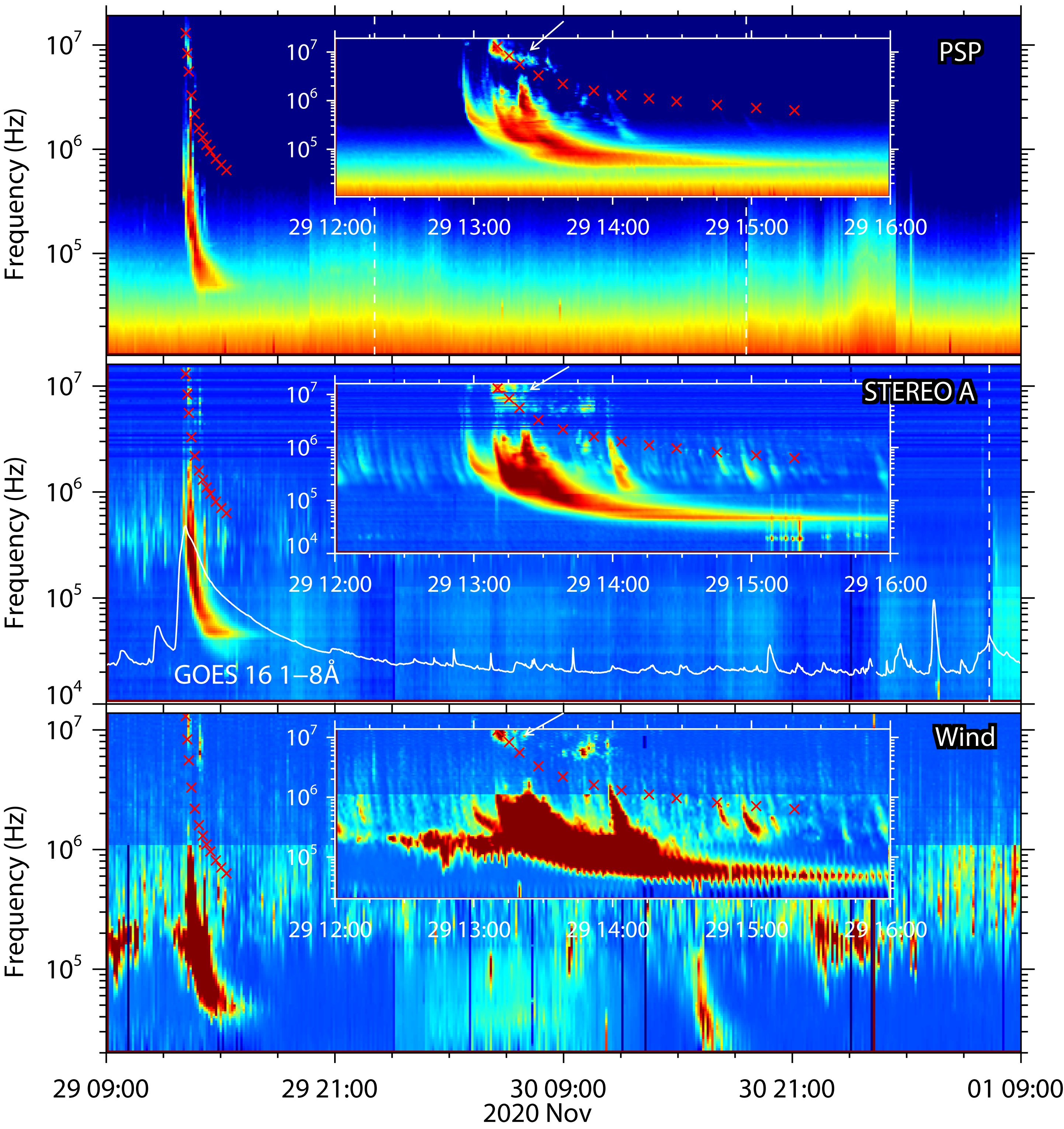}
	\caption{\label{f7}Radio dynamic spectra associated with the 2020 November 29 eruption from PSP, STEREO A, and Wind. The white arrows mark the type II radio burst observed by the spacecraft. The shock nose distances derived from the ellipsoid model are converted to frequencies by using the Leblanc density model with an electron density of 15 cm$^{-3}$ at 1 AU. The frequencies are overlapped in the spectra as red crosses. The white curve in the middle panel is the GOES X--ray flux scaled in arbitrary units. The vertical dashed lines represent the shock arrival times from in situ measurements at PSP and STEREO A. The areas with radio bursts are expanded and plotted over each image for clarity.} 
\end{figure}

Figure \ref{f7} shows the radio dynamic spectra associated with the 2020 November 29 CME from PSP, STEREO A, and Wind. The intense type III radio busts, which start at about 12:55 UT on November 29 within several minutes after the liftoff of the CME, are observed by all three spacecraft. An intense type III radio bust portends the occurrence of a major CME on the Sun (e.g., \citealt{Reiner2001}). All the spacecraft observe a short--duration type II radio burst in the initial phase, which starts at about 13:10 UT almost near the peak of the X--ray flux, as shown in the panels zoomed in. We use the Leblanc density model \citep{Leblanc1998} with an electron density of 15 cm$^{-3}$ at 1 AU to convert the shock nose distances obtained from the ellipsoid model to frequencies. The Leblanc density model describes the average radial variation of the density of the medium where the shock was propagating, so the electron density of 15 cm$^{-3}$ is the nominal density and not necessary to be observed at 1 AU. The frequencies are doubled to their harmonic frequencies and overlapped on the spectra. The frequencies roughly match the observed type II radio burst when the corresponding shock nose distances are below $\sim$ 5 \rsun{}. The vertical dashed lines in the PSP spectrum represent the shock arrival times at PSP from in situ measurements. The previous one is associated with the small CME on 2020 November 26. The second shock is the one we study in this paper, and is also observed at STEREO A. The arrival times of the second shock at PSP and STEREO A correspond to the suddenly enhanced, diffuse intensity in the spectra. Note that there is no shock signature observed at Wind.  

\begin{figure}
	\epsscale{1.2}
	\plotone{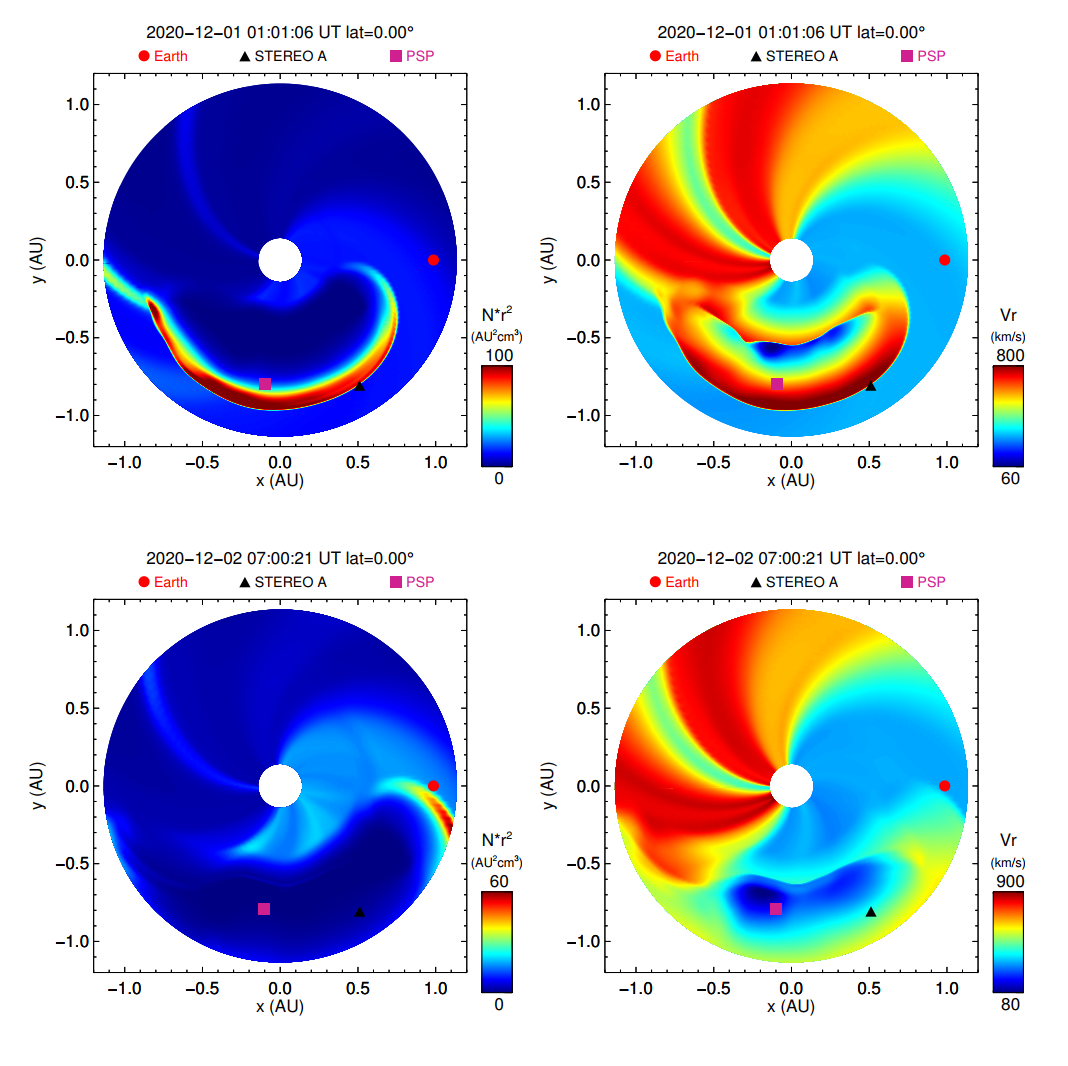}
	\caption{\label{f8}Density and radial speed distributions from the WSA--ENLIL simulation in the ecliptic plane at 01:01:06 UT on 2020 December 1 (top) and 07:00:21 UT on 2020 December 2 (bottom). The values of the density and speed are scaled by the color bar at the right corner of each panel. The positions of the Earth, STEREO A, and PSP are marked by the circle, triangle, and square, respectively. The first time indicates the moment when the CME have just passed PSP and would arrive at STEREO A. The second time is close to the CME arrival time at the Earth.} 
\end{figure}

To determine the propagation and heliospheric impacts of the CME/shock, we request a WSA--ENLIL simulation \citep{Arge2004, Odstrcil2004} run from the Community Coordinated Modeling Center (CCMC). CME parameters derived from the GCS model, such as the time across the inner boundary at 21.5 \rsun{}, half angular width, radial speed, and propagation direction, are input in the simulation. Note that, as mentioned above, a small CME occurred on 2020 November 26 and was also observed by PSP. Here, we only insert the CME of interest in the simulation, and the results are good compared with the in situ measurements. The CME/shock arrival times at PSP and Wind predicted by the WSA–ENLIL model are consistent with the observed arrival times from in situ measurements. As for STEREO A, the predicted arrival time is only about 5 hours earlier than the observed shock arrival time. The density and velocity profiles from the simulation generally agree with the in situ measurements, although the simulation overestimates the density. The simulation results are displayed in Figure \ref{f8}. The top panels show the density and radial speed distributions at 01:01:06 UT on 2020 December 1. The CME arrives at PSP and STEREO A with a large angle width. The density and radial speed of the CME are much larger than the ambient solar wind. From the bottom panels, the far flank of the CME (or the CME leg) could arrive at the Earth with relatively low density and speed about one day after arriving at STEREO A. 

\section{in situ measurements} \label{insitu}
\begin{figure}
	\epsscale{1.2}
	\plotone{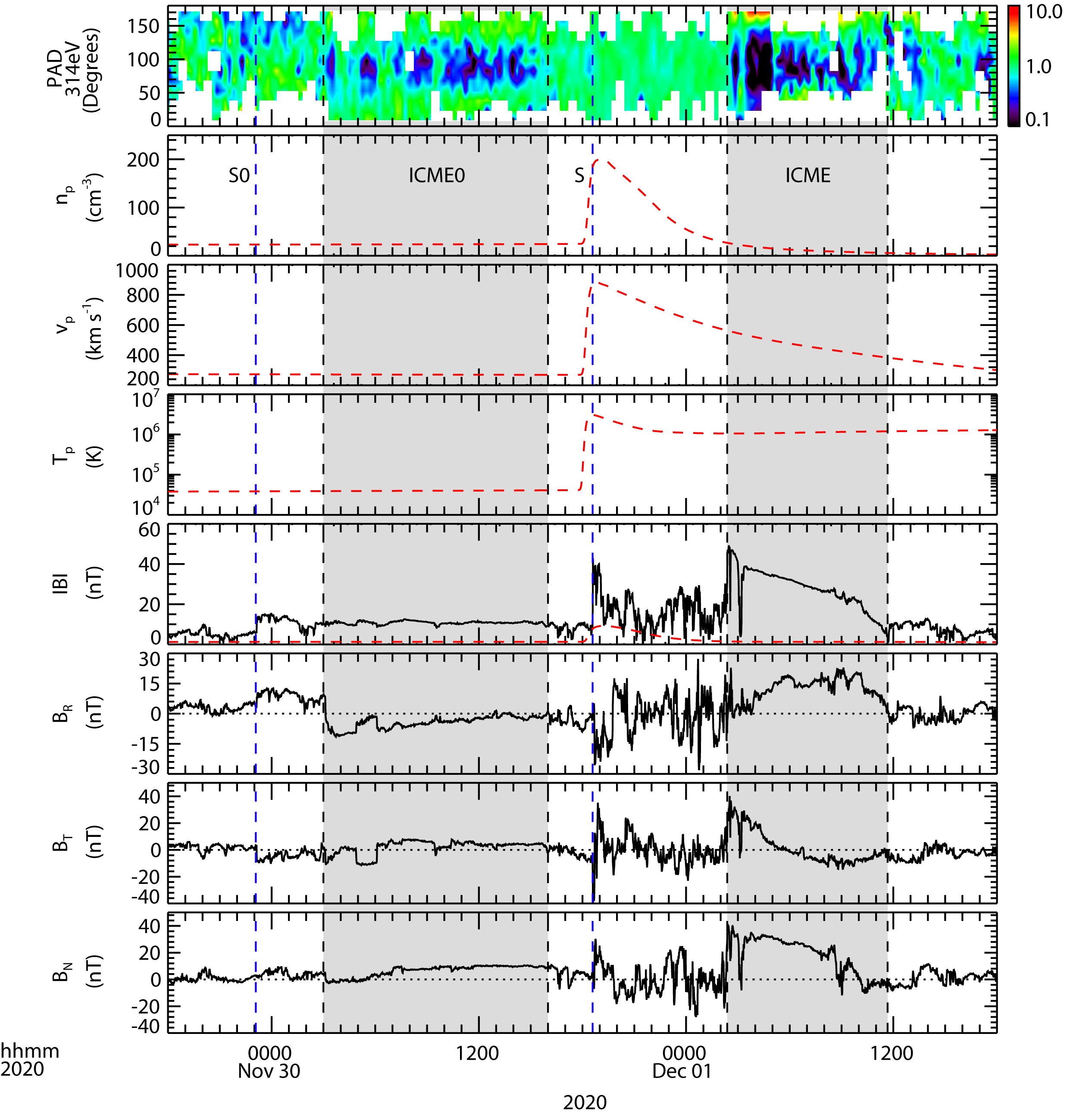}
	\caption{\label{f9}In situ solar wind measurements and simulation results at PSP. From top to bottom, the panels show the normalized 314 eV electron PAD, proton density, bulk speed, proton temperature, magnetic field strength, and components, respectively. Note that there are no plasma data from PSP during the time of interest. The red dashed lines represent the simulation results. The blue vertical dashed lines indicate the observed arrival times of the shocks marked with ``S0'' and ``S'', respectively. The shaded regions indicate the intervals of two ICMEs.} 
\end{figure}

Figure \ref{f9} displays the solar wind magnetic field measurements from the FIELDS instruments \citep{Bale2016} aboard PSP, and the WSA--ENLIL simulation results at PSP. From PSP in situ measurements, there are two shocks and two ICME structures observed at PSP. The first shock passes PSP around 23:05 UT on November 29, which may be associated with the small CME event on November 26 propagating toward PSP. The second shock is the one of interest, which is observed at PSP around 18:35 UT on November 30, followed by a long--duration sheath region and an ICME structure. The long--duration sheath region is from 18:35 UT on November 30 to 02:23 UT on December 1 lasting about 8 hours. This indicates that PSP likely observes the flank of the CME, because the standoff distance between the shock front and the CME flux rope increases from the nose to the flank (see Figure \ref{f4}). The propagation direction of the CME in latitude and the location of the source region also imply that the northern flank of the CME would arrive at PSP and STEREO A. The ICME interval is from 02:23 UT to 11:40 UT on December 1, which is determined by the bidirection of the electron pitch angle distribution (PAD) and magnetic field profiles, such as the smooth and strong magnetic field and the coherent rotation of the field. Although the resolution of the electron PAD is low, there are some indications of the bidirectional streaming electron (BDE) signatures. We can see the mainly northward $B_{N}$ component in the ICME implying a large tilt angle of the CME flux rope. This is consistent with the results obtained from the GCS model. The WSA--ENLIL simulation results are plotted as red dashed lines. Because there are no plasma data from PSP during the period of time, we can not compare the simulation results with the in situ plasma measurements. The shock arrival time predicted by the WSA--ENLIL model is consistent with the observed shock arrival time. Therefore, we use the WSA--ENLIL simulation results to estimate the shock speed at PSP, which is about 850 \kms{}. The simulated magnetic field is much lower than the observed, because in the WSA--ENLIL model, the inserted CME has no internal magnetic field and carries the field from the ambient solar wind. From the in situ measurements at PSP, the second shock has not propagated into the first ICME yet.

\begin{figure}
	\epsscale{1.2}
	\plotone{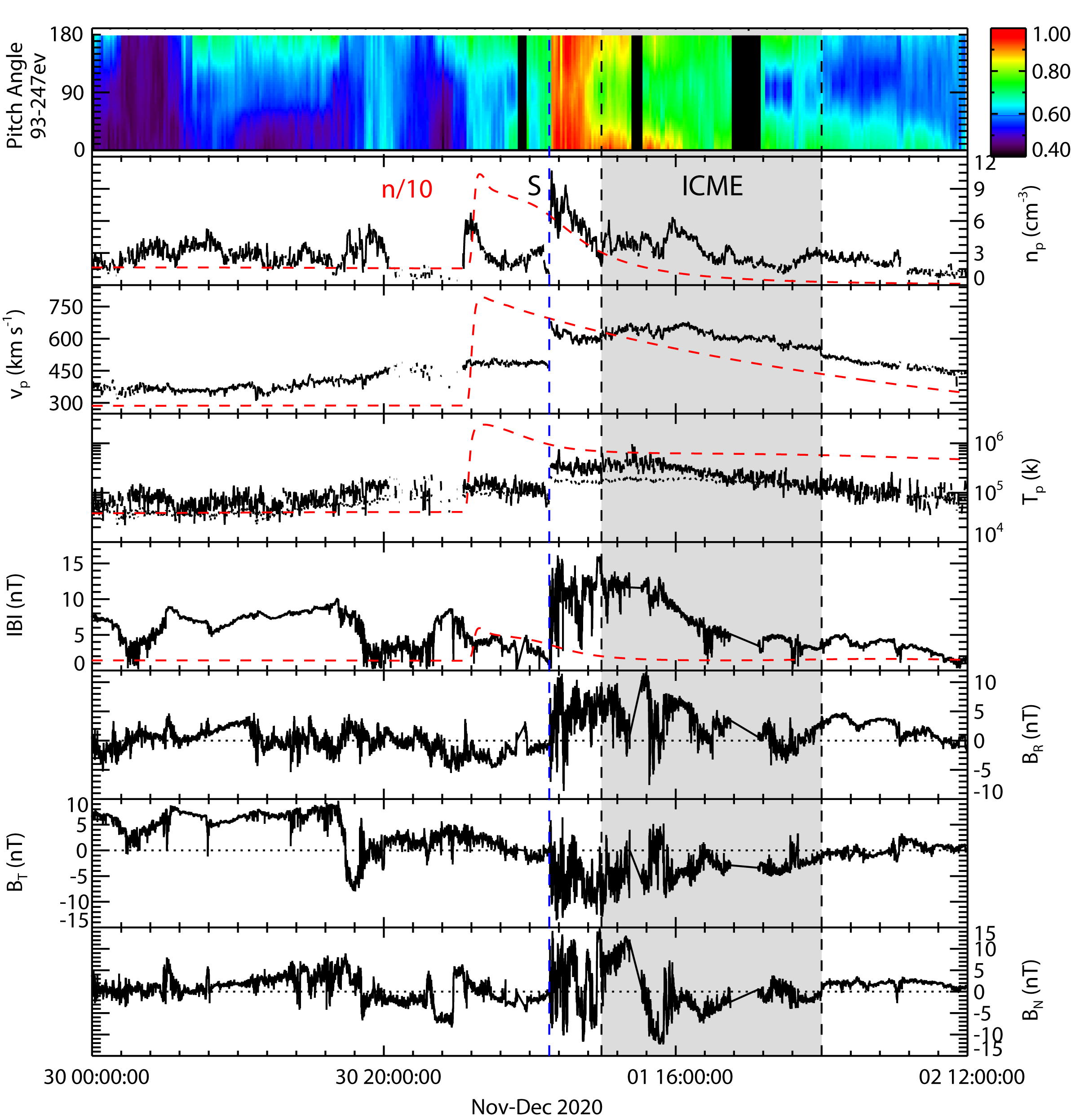}
	\caption{\label{f10}In situ solar wind measurements and simulation results at STEREO A. From top to bottom, the panels show the normalized 93--247 eV electron PAD, proton density, bulk speed, proton temperature, magnetic field strength, and components, respectively. The red dashed lines represent the simulation results, and the density from the simulation  is divided by a factor of 10. The dotted line in the fourth panel denotes the expected proton temperature calculated from the observed speed \citep{Lopez1987}. The blue vertical dashed line indicates the observed arrival time of the shock marked with ``S'', and the shaded region indicates the interval of the ICME.} 
\end{figure}

The in situ measurements from STEREO A associated with the 2020 November 29 CME are presented in Figure \ref{f10}, with the simulation results at STEREO A shown as red dashed lines. A shock is observed at STEREO A around 07:20 UT on December 1, which is $\sim$5 hours later than the predicted arrival time from the simulation. The shock speed predicted by the WSA--ENLIL model is about 850 \kms{}, which is larger than the observed shock speed at STEREO A (about 700 \kms{}). The shock is followed by a sheath region lasting about 4 hours, and the ICME interval is determined by the electron PAD and magnetic field profiles. The magnetic field strength of the ICME from STEREO A shows a declining profile similar to the observations from PSP. However, the magnetic field components are more dynamic than that observed by PSP. 

\begin{figure}
	\epsscale{1.2}
	\plotone{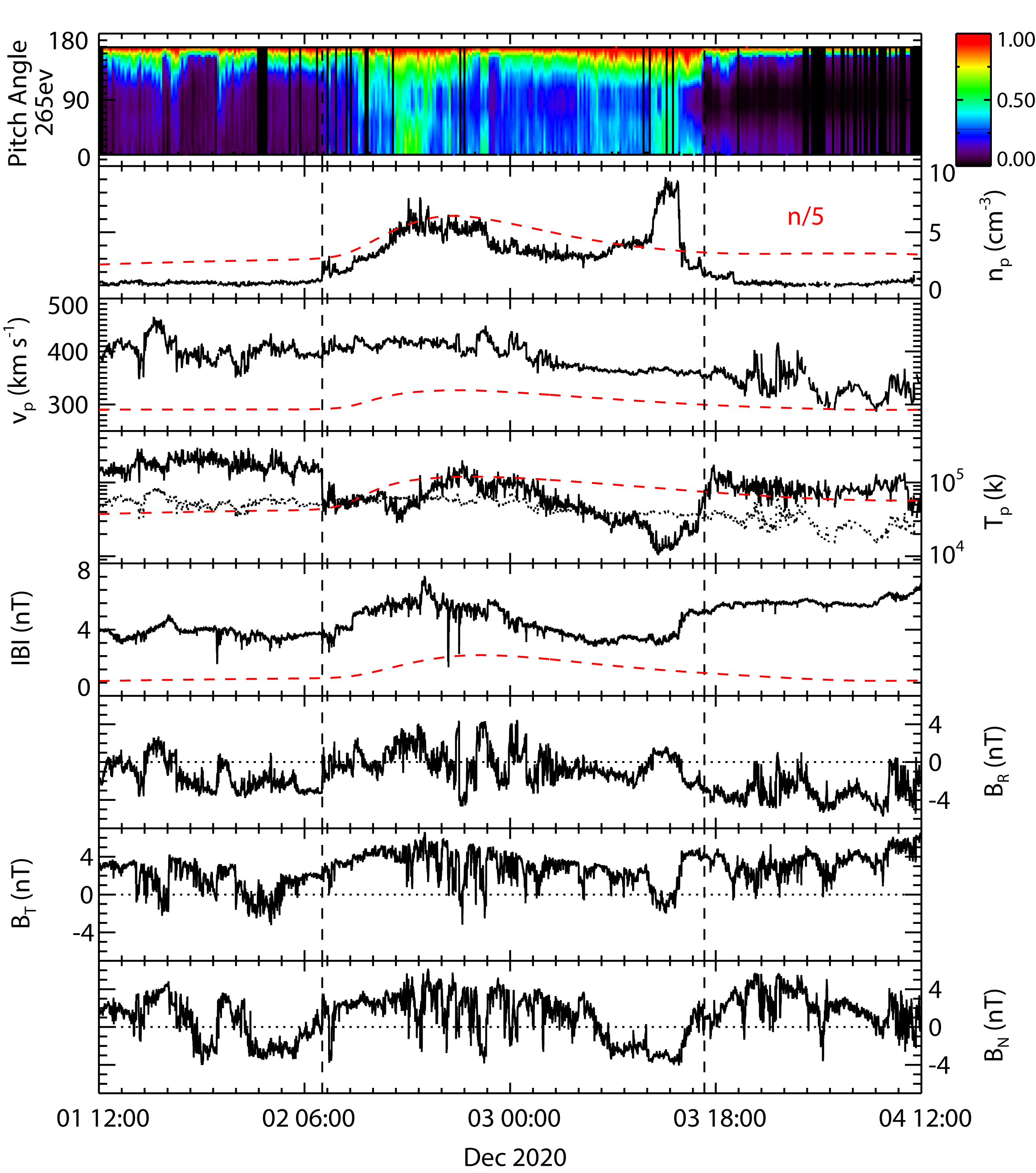}
	\caption{\label{f11}In situ solar wind measurements and simulation results at Wind. From top to bottom, the panels show the normalized 265 eV electron PAD, proton density, bulk speed, proton temperature, magnetic field strength, and components, respectively. The red dashed lines represent the simulation results, and the density from the simulation is divided by a factor of 5. The dotted line in the fourth panel denotes the expected proton temperature calculated from the observed speed \citep{Lopez1987}. The black vertical dashed lines indicate the interval of the ICME--like structure.} 
\end{figure}

If we only consider the location of the source region and the propagation direction of the CME/shock, the CME and shock may not arrive at the Earth, because the Earth is about 90\degr{} away from the propagation direction of the CME/shock. Figure \ref{f11} shows the Wind in situ measurements probably associated with the CME, which seem to be not like a typical ICME (e.g., \citealt{ZurbuchenT2006}). Given the large expansion of the CME and shock near the Sun, and the simulation suggests that the CME may arrive at the Earth (see Figure \ref{f8}), we infer that the structure observed at Wind may be a part of the CME. The lack of typical ICME signatures (e.g., no clear BDE signature) probably suggests that the far flank of the CME (or the CME leg) arrives at the Earth. According to the proton temperature and magnetic field strength profiles, we give an approximate interval of the ICME, as indicated by two vertical dashed lines. The density profile from the simulation generally agrees with the in situ measurements, although the WSA--ENLIL simulation overestimates the density about 5 times. The CME arrival time (07:33 UT on December 2) predicted by the WSA--ENLIL model is also in good agreement with the measurements. These indicate that the CME indeed arrives at the Earth. According to the CME and shock structures in the coronagraph images and the simulation results, we suggest that the far flank of the CME (or the CME leg) arrives at the Earth due to the vast expansion of the CME. However, there is no shock signature in the measurements. This supports the perspective of \citet{Liu2017} that, at some point, the shock would be just a wave and lose the nonlinear steepening character especially near the wake.

\section{summary and conclusions} \label{sum}
We have investigated the source eruption, propagation and expansion characteristics, and heliospheric impacts of the 2020 November 29 CME/shock, combining multi--perspective remote sensing observations and multipoint in situ measurements. EUV observations together with PFSS modeling are used to examine the source eruption signatures and the coronal magnetic field configuration. The structures and kinematics of the CME and associated shock are analyzed using the forward models and remote sensing observations from multiple spacecraft. The WSA--ENLIL simulation and the multipoint in situ measurements are used to study the heliospheric impacts of the CME/shock.
Key results are obtained concerning the source eruption, interplanetary propagation, and heliospheric impacts of the CME and associated shock.

The source region is beneath toroidal closed magnetic field arches, which are surrounded by open magnetic field lines of the same polarity. The CME eruption is associated with a significant dimming and an EUV wave sweeping a large area of the solar disk. This indicates a large CME eruption associated with a shock, and implies a rapid expansion of the CME/shock. This is a limb event as viewed from the Earth, and the CME flux rope can be seen in AIA 131 \AA{} images, whose axis has a large tilt angle. Therefore, we can track the flux rope at its early stage using EUV observations.

Due to the large expansion of the CME flux rope and its fast acceleration, the shock shows an ellipsoidal structure, expands in all directions, and encloses the whole Sun as viewed from both SOHO and STEREO A. We use the GCS model and the ellipsoid model to fit the CME and the shock, respectively. The tilt angle and propagation direction of the CME flux rope obtained from the GCS model are consistent with the EUV observations. The CME flux rope quickly expands and accelerates at its initial stage. Therefore, the ellipsoidal structure of the shock is produced by the large expansion of the CME flux rope and its fast acceleration. The ellipsoid model fitting results suggest that the shock expands in all directions and encloses the whole Sun as viewed from both SOHO and STEREO A, which helps to explain the detection of SEPs at Solar Orbiter behind the location of the eruption. The structure and kinematics of a shock can be used to study the relationship between SEPs arriving at different locations and their sources (e.g., \citealt{Rouillard2011, Lario2017, Zhu2021}). The study of the 2020 November 29 widespread SEP event has been performed by \citet{Kollhoff2021}, but they did not give a detailed analysis of the CME and associated shock kinematics. Therefore, the shock ellipsoid fitting results (see Table \ref{tab:ellipsoid_fitting}) in our paper can be a complement and provide some key parameters for the study of the SEP acceleration.

The structure and potential impacts of the shock are mainly determined by its radial and lateral expansions. We track the CME by producing jmaps with projection effects minimized, as the propagation direction is almost perpendicular to the Sun--Earth line. The kinematics of CME/shock from different methods agree with each other. The speeds of the CME and shock nose are rapidly accelerated to maxima within only 1 hour below $\sim$6 \rsun{}, and then show a deceleration, which is a typical speed profile for fast CMEs \citep{Liu2013}. The shock expansion speeds are much larger than the translation speed, so we suggest that the structure and potential impacts of the shock are mainly determined by its radial and lateral expansions. The liftoff of the CME flux rope and the rise of the X--ray flux are almost at the same time, and all the speeds of the CME and shock increase during the flare rising phase and peak after the maximum value of the X--ray flux. 

The type III and type II radio bursts associated with the eruption are observed at PSP, STEREO A, and Wind. All three spacecraft observe an intense type III radio burst, which indicates a major CME on the Sun, and a short--duration type II radio burst in the initial phase. We convert the shock nose distances obtained from the ellipsoid model to frequencies using the Leblanc density model with an electron density of 15 cm$^{-3}$ at 1 AU. The shock kinematics seem consistent with the frequency drift of the associated type II burst. 
 
According to the separated in situ measurements and simulation results, the CME and shock arrive at PSP and STEREO A, and the Earth observe the far flank of the CME (or the CME leg) due to the large expansion of the CME. The shock is observed at PSP around 18:35 UT on November 30, followed by a long--duration sheath region and an ICME structure. The ICME has a mainly northward $B_{N}$ component implying a large tilt angle of the CME flux rope, which is consistent with the results obtained from the GCS model. The shock passes STEREO A around 07:20 UT on December 1. The magnetic field strength of the ICME from STEREO A shows a declining profile similar to the observations from PSP, but the magnetic field components are more dynamic than that observed by PSP. The Earth is about 90\degr{} away from the CME/shock nose, so according to the location of the source region and the propagation direction of the CME and shock, they may not arrive at the Earth. Moreover, it is difficult to predict whether and when the CME/shock would arrive at the Earth only based on the remote sensing observations. Combining the in situ measurements from Wind and WSA--ENLIL simulation results, we finally find that, because of the large expansion of the CME, the far flank of the CME (or the CME leg) arrives at the Earth with no shock signature. This is surprising because the shock is generally thought to be wider than the CME. The shock may have already decayed before reaching Wind along this far flank direction. These results highlight that multispacecraft remote sensing and in situ observations are important for determining the heliospheric impacts of CMEs.

\acknowledgments
The research was supported by NSFC under grant 41774179, Beijing Municipal Science and Technology Commission (Z191100004319003), the Specialized Research Fund for State Key Laboratories of China, and the Strategic Priority Research Program of Chinese Academy of Sciences (XDA15018500). We acknowledge the use of data from Parker Solar Probe, \textit{STEREO}, \textit{SDO}, \textit{SOHO}, \textit{GOES}, and \textit{Wind}. The WSA--ENLIL simulations are provided by CCMC through their public Runs on Request system (\url{http//ccmc.gsfc.nasa.gov}).

\clearpage

\clearpage

\clearpage

\clearpage

\clearpage

\clearpage

\clearpage

\clearpage


\clearpage

\clearpage

\clearpage

\clearpage
\begin{deluxetable*}{ccccccc}
	\tablecaption{The shock ellipsoid fitting parameters and corresponding times.\label{tab:ellipsoid_fitting}}
	\tablewidth{0pt}
	\tablehead{
		\multicolumn2c{Time (UT)} & \colhead{$\theta$}  & \colhead{$\phi$} &
		\colhead{$h$} & \colhead{$a$} & \colhead{$c$} \\
		\colhead{STEREO A} & \colhead{GOES/SOHO}  & \colhead{(\degr{})} & \colhead{(\degr{})} &
		\colhead{\rsun{}} & \colhead{\rsun{}} & \colhead{\rsun{}}
	}
	\startdata
	EUVI 12:55:30 & SUVI 12:56:37 & -95	 & -18 & 1.31 & 0.51 & 0.51 \\
	COR1 13:10:18 & SUVI 13:10:37 & -90	 & -15 & 1.79 & 1.27 & 1.06 \\
	COR1 13:15:18 & SUVI 13:15:27 & -90	 & -15 & 2.00 & 1.52 & 1.35 \\
	COR1 13:20:18 & SUVI 13:20:37 & -85	 & -15 & 2.34 & 2.12 & 1.83 \\
	COR2 13:24:00 & C2 13:23:38	 & -80	 & -15 & 2.73 & 3.10 & 2.38 \\
	COR2 13:39:00 & C3 13:39:54	 & -80	 & -23 & 3.16 & 5.49 & 4.77 \\
	COR2 13:54:00 & C3 13:51:55	 & -80	 & -25 & 3.52 & 7.65 & 6.51 \\
	 -- & C3 14:03:54 & -80	 & -25 & 3.83 & 9.43 & 8.21 \\
	 -- & C3 14:15:38 & -80	 & -27 & 4.10 & 11.34 & 9.84 \\
	COR2 14:24:00 & C3 14:27:39  & -80	 & -30 & 4.41 & 13.10 & 11.53 \\
	COR2 14:39:00 & C3 14:39:38  & -80	 & -30 & 4.68 & 14.85 & 13.16 \\
	 -- & C3 15:07:59  & -80	 & -33 & 5.49 & 18.53 & 16.90 \\
	 -- & C3 15:18:34  & -80	 & -35 & 5.71 & 19.88 & 18.24 \\ 
	\enddata
	\tablecomments{$\theta$, $\phi$ and $ h $ are the longitude, latitude, and height of the center of the ellipsoid, respectively. $a$ and $c$ are the lengths of the two semiprincipal axes. Because we assume the cross section of the shock ellipsoid perpendicular to the propagation direction to be a circle, there are only 5 free parameters in our fitting.}
\end{deluxetable*}

\clearpage
\bibliography{article}
\bibliographystyle{aasjournal}

\end{document}